\documentclass[twocolumn,11pt]{article}
\usepackage{times}
\usepackage{graphicx}
\newcommand{\R}{{\rm I}\!{\rm R}} 

\begin{document}
\global\def\refname{{\normalsize \it References:}}
\baselineskip 12.5pt
%
%
%
\title{\LARGE \bf Statistical Complexity in Traveling Densities}

\date{}

\author{\hspace*{-10pt}
\begin{minipage}[t]{2.7in} \normalsize \baselineskip 12.5pt
\centerline{RICARDO L\'OPEZ-RUIZ}
\centerline{Universidad de Zaragoza}
\centerline{Department of Computer Science and BIFI}
\centerline{Campus San Francisco, E-50009 Zaragoza}
\centerline{SPAIN}
\centerline{rilopez@unizar.es}
\end{minipage} \kern 0in
\begin{minipage}[t]{2.7in} \normalsize \baselineskip 12.5pt
\centerline{JAIME SA\~NUDO}
\centerline{Universidad de Extremadura}
\centerline{Department of Physics and BIFI}
\centerline{Avda. de Elvas, E-06071 Badajoz}
\centerline{SPAIN}
\centerline{jsr@unex.es}
\end{minipage}
%
%
\\ \\ \hspace*{-10pt}
\begin{minipage}[b]{6.9in} \normalsize
\baselineskip 12.5pt {\it Abstract:}
{\small In this work, we analyze the behavior of statistical complexity in several systems 
where two identical densities that travel in opposite direction cross each other.
The crossing between two Gaussian, rectangular and triangular densities is studied in detail.
For these three cases, the shape of the total density presenting an extreme value in complexity 
is found.} 
\\ [4mm] {\it Key--Words:}
Statistical complexity, traveling densities, distribution shape
\end{minipage}
\vspace{-10pt}}

\maketitle

\thispagestyle{empty} \pagestyle{empty}
%
%
\section{Introduction}
\label{S1} \vspace{-4pt}

An interesting problem that has not been broadly investigated
in the literature is the behavior of the statistical complexity
in time-dependent systems. A work in this direction was done in
Ref. \cite{calbet2001} where this behavior was studied in a gas
out of equilibrium decaying toward the asymptotic equilibrium state.

In this communication, we carry out the study of the statistical 
complexity $C$ in a time-dependent system $\rho(x,t)$ composed of 
two one-dimensional (variable $x$) identical densities that travel 
in opposite directions with the same velocity $v$, one of them, 
$\rho_+(x,t)$, going to the right and the other one, $\rho_-(x,t)$ 
going to the left. That is  
\begin{equation}
\rho(x,t)={1\over 2}\,\rho_+(x,t)+{1\over 2}\,\rho_-(x,t),
\label{eq-rho}
\end{equation}
with the normalization condition $\int_{\R}\rho_{\pm}(x,t)dx=1$
that implies the normalization of $\rho(x,t)$.
In the next section, we perform the analysis of $C$ for two Gaussian, rectangular and 
triangular traveling densities, which verify the initial condition 
$\rho_+(x,0)=\rho_-(x,0)$. Specifically, the shape of $\rho(x,t)$ presenting
the maximum and minimum $C$ is found for these three cases. The final section
includes our conclusions.

\section{Complexity in Traveling Densities}
\label{S2} \vspace{-4pt}

Let us start by recalling the definition of the statistical complexity $C$ \cite{lopez1995},
the so-called $LMC$ complexity, that is defined as
\begin{equation}
C = H\cdot D\;,
\end{equation}
where $H$ represents the information content of the system and $D$ gives an idea
of how much concentrated is its spatial distribution. 
For our purpose, we take a version used in Ref. \cite{lopez2002}
as quantifier of $H$. This is the simple exponential Shannon entropy \cite{dembo1991},
that takes the form, 
\begin{equation}
H = e^{S}\;,
\end{equation}
where $S$ is the Shannon information entropy \cite{shannon1948},
\begin{equation}
S = -\int p(x)\;\log p(x)\; dx \;,
\label{eq1}
\end{equation}
with $x$ representing the continuum of the system states and $p(x)$
the probability density associated to all those states.
We keep for the disequilibrium the form originally introduced in 
Refs. \cite{lopez1995,lopez2002}, that is,
\begin{equation}
D = \int p^2(x)\; dx\;.
\label{eq2} 
\end{equation}
Now we proceed to calculate $C$ for the system above mentioned (\ref{eq-rho})
in the Gaussian, rectangular and triangular cases. 

\subsection{Gaussian case}

Here the two one-dimensional traveling densities that compose system (\ref{eq-rho}) 
take the form: 
\begin{equation}
\rho_{\pm}(x,t)={1\over \sigma\sqrt{2\pi}}\,\exp\left\{{-{(x\mp vt)^2}\over 2\sigma^2}\right\},
\label{eq-gauss}
\end{equation}
where $\sigma$ is the variance of the density distribution.

\begin{figure}[t]  
\centerline{\includegraphics[width=9cm]{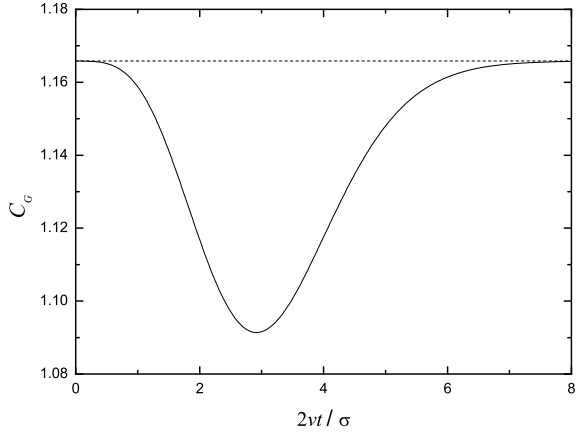}}     
\caption{\small Statistical complexity, $C_G$, vs. the adimensional separation, $2vt/\sigma$, 
between the two traveling Gaussian densities defined in Eq. (\ref{eq-gauss}).
The minimum of $C_G$ is reached when $2vt/\sigma=2.91$.
The dashed line indicates the value of complexity for the normalized Gaussian distribution.}  
\label{fig1}  
\end{figure}

\begin{figure}[t]  
\centerline{\includegraphics[width=9cm]{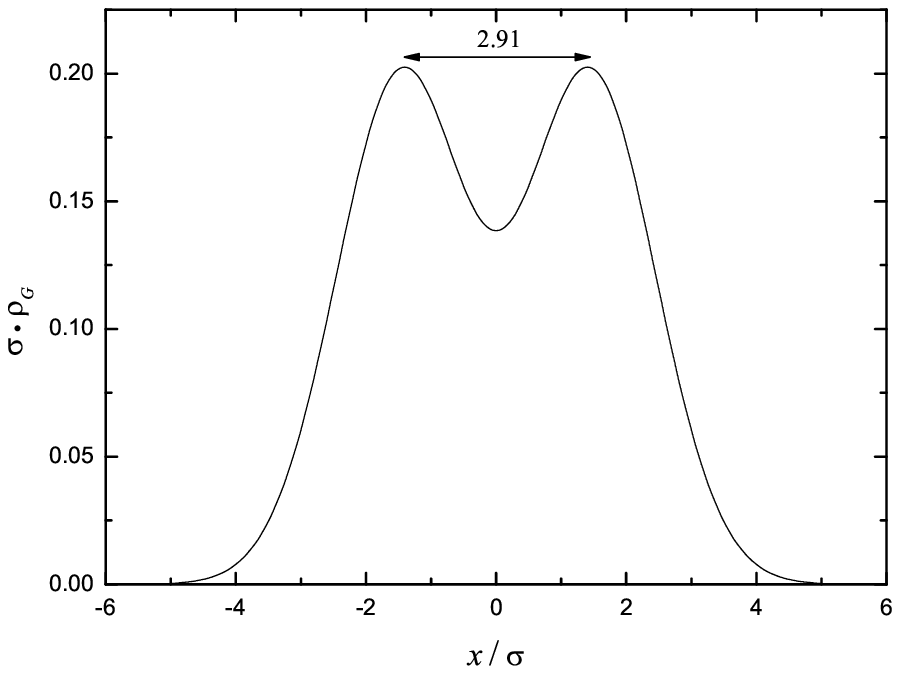}}     
\caption{\small Shape of the density (\ref{eq-rho}) in adimensional units that presents the 
minimum statistical complexity when the two traveling Gaussian densities defined in (\ref{eq-gauss}) 
are crossing. Notice that the value of the adimensional
separation between the centers of both Gaussian distributions must be $2.91$.}  
\label{fig2}  
\end{figure}

The behavior of complexity, $C_G$, as a function of the adimensional quantity $2vt/\sigma$ is given 
in Fig. \ref{fig1}. Let us observe that $C_G$ presents a minimum. The shape of system (\ref{eq-rho})
for this minimum complexity case is plotted in an adimensional scale in Fig. \ref{fig2}.

\subsection{Rectangular case}

Now the two one-dimensional traveling densities that compose system (\ref{eq-rho}) 
take the form: 
\begin{equation}
\rho_{\pm}(x,t)=\left\{
\begin{array}{llc}
1/\delta & \mbox{if} & -\delta/2\leq x\mp vt\leq\delta/2\,, \\
& \\
0 & \mbox{if} & |x\mp vt|>\delta/2\,.
\end{array}
\right.
\label{eq-rectang}
\end{equation}
where $\delta$ is the width of each distribution.

For this case, the complexity, $C_R$, can be analytically obtained. Its expression is:
\begin{equation}
C_R(t)=\left\{
\begin{array}{llc}
2^{2vt/\delta}\left(1-{vt\over \delta}\right)& \mbox{if} & 0\leq 2vt \leq\delta\,, \\
& \\
1 & \mbox{if} & 2vt >\delta\,,
\end{array}
\right.
\label{eq-C-rect}
\end{equation}

\begin{figure}[h]  
\centerline{\includegraphics[width=9cm]{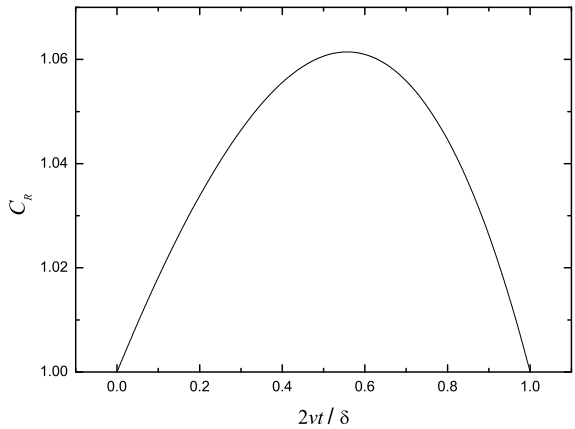}}     
\caption{\small Statistical complexity, $C_R$, vs. the adimensional separation, $2vt/\delta$, 
between the two traveling rectangular densities defined in Eq. (\ref{eq-rectang}).
The maximum of $C_R$ is reached when $2vt/\delta=0.557$.
Observe that the normalized rectangular distribution has $C_R=1$.}  
\label{fig3}  
\end{figure}  

\begin{figure}[h]  
\centerline{\includegraphics[width=9cm]{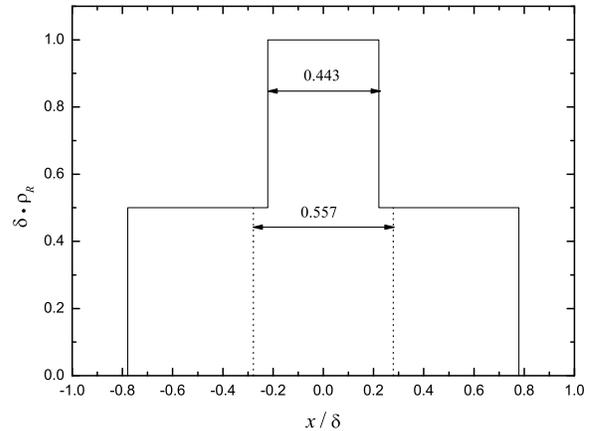}}     
\caption{\small Shape of the density (\ref{eq-rho}) in adimensional units that presents the maximum statistical complexity
when the two traveling rectangular densities defined in (\ref{eq-rectang}) are crossing. 
Notice that the value of the adimensional separation between the centers of both rectangular distributions must 
be $0.557$. Then, the width of the overlapping between both distributions is 0.443.}  
\label{fig4}  
\end{figure}  

The behavior of $C_R$ as a function of the adimensional quantity $2vt/\delta$ is given 
in Fig. \ref{fig3}. Let us observe that $C_R$ presents a maximum. The shape of system (\ref{eq-rho})
for this maximun complexity case is plotted in an adimensional scale in Fig. \ref{fig4}.

\subsection{Triangular case}

The two one-dimensional traveling densities that compose system (\ref{eq-rho}) 
take the form in this case: 
\begin{equation}
\rho_{\pm}(x,t)=\left\{
\begin{array}{llc}
{(x\mp vt)\over\epsilon^2} + {1\over\epsilon} & \mbox{if} & -\epsilon\leq x\mp vt\leq 0 \,, \\
& & \\
{-(x\mp vt)\over\epsilon^2} + {1\over\epsilon} & \mbox{if} & 0 < x\mp vt\leq\epsilon \,, \\
& & \\
0 & \mbox{if} & |x\mp vt|>\epsilon \,,
\end{array}
\right.
\label{eq-triang}
\end{equation}
where $\epsilon$ is the width of each distribution (isosceles triangle whose base length is $2\epsilon$ ).

\begin{figure}[h]  
\centerline{\includegraphics[width=9cm]{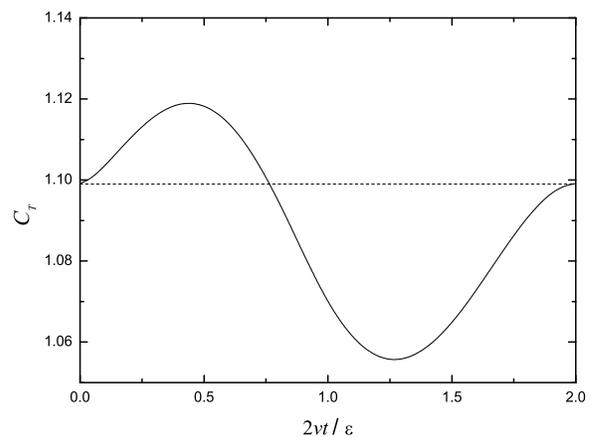}}     
\caption{\small Statistical complexity, $C_T$, vs. the adimensional separation, $2vt/\epsilon$, 
between the two traveling triangular densities given in Eq. (\ref{eq-triang}).
The maximum and minimum of $C_T$ are reached when $2vt/\epsilon$ takes the values $0.44$ and $1.27$,
respectively. The dashed line indicates the value of complexity for the normalized triangular distribution.}  
\label{fig5}  
\end{figure}

\begin{figure}[h]  
\centerline{\includegraphics[width=9cm]{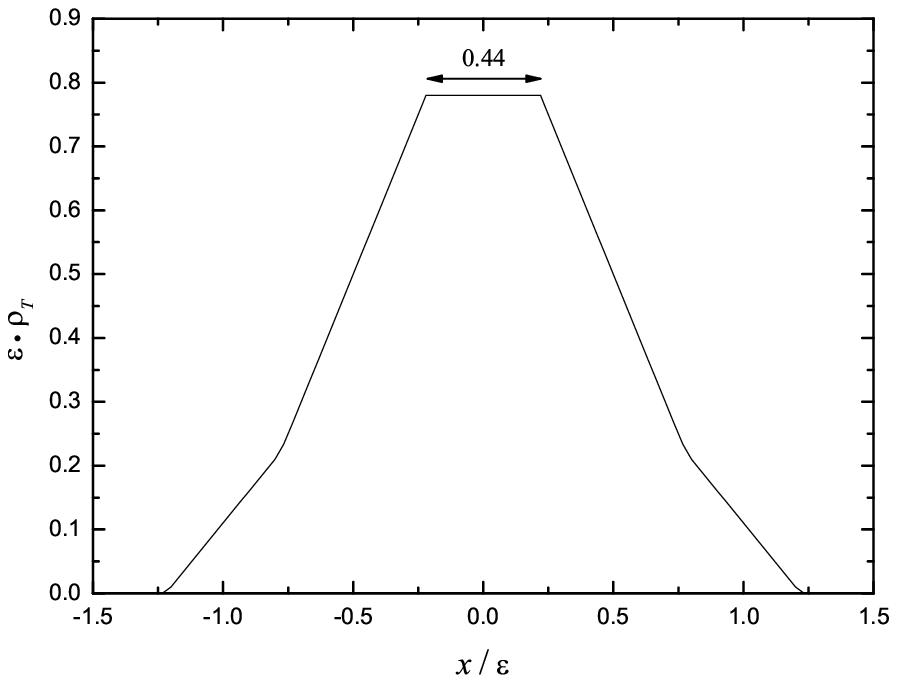}}     
\caption{\small Shape of the density (\ref{eq-rho}) in adimensional units that presents 
the maximum statistical complexity
when the two traveling triangular densities defined in (\ref{eq-triang}) are crossing. 
Notice that the value of the adimensional separation between the centers of both triangular 
distributions must be $0.44$.}  
\label{fig6}  
\end{figure}

\begin{figure}[h]  
\centerline{\includegraphics[width=9cm]{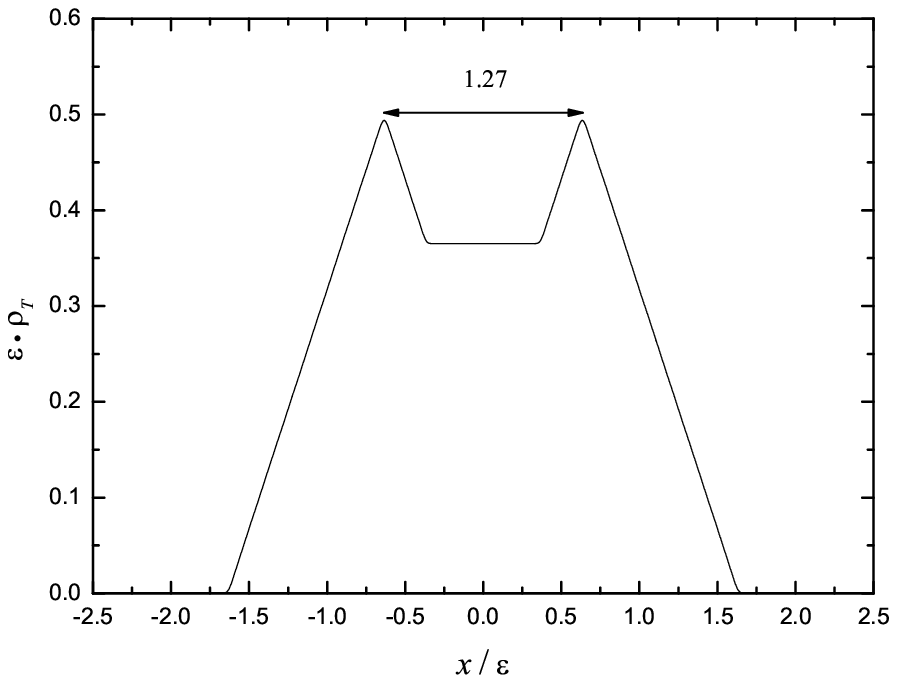}}     
\caption{\small Shape of the density (\ref{eq-rho}) in adimensional units that presents 
the minimum statistical complexity
when the two traveling triangular densities defined in (\ref{eq-triang}) are crossing. 
Notice that the value of the adimensional separation between the centers of both triangular 
distributions must be $1.27$.}  
\label{fig7}  
\end{figure}

The behavior of complexity, $C_T$, as a function of the adimensional quantity $2vt/\epsilon$ is given 
in Fig. \ref{fig5}. Let us observe that $C_T$ presents a maximum and a minimum. The shape of system (\ref{eq-rho})
for both cases, with maximum and minimum complexitiy, are plotted in an adimensional scale in 
Figs. \ref{fig6} and \ref{fig7}, respectively.

\section{Conclusion}
\label{S3} \vspace{-4pt}

In this communication, we have studied the behavior of the statistical complexity
as a function of time when two traveling identical densities are crossing each other.
Three cases have been analyzed: Gaussian, rectangular and triangular densities.
The Gaussian case presents a configuration with minimum complexity. The rectangular case
displays a configuration with maximum complexity. The triangular case shows an intermediate
behavior between the two former cases with a maximum complexity configuration and another one
with minimum complexity.

\vspace{10pt} \noindent
{\bf Acknowledgements:} This research was supported by the
spanish Grant with Ref. FIS2009-13364-C02-C01.
J.S. also thanks to the Consejer\'ia de Econom\'ia, Comercio e Innovaci\'on 
of the Junta de Extremadura (Spain) for financial support, Project Ref. GRU09011.

\end{document}